# UTILIZING MACHINE LEARNING TO PREVENT WATER MAIN BREAKS BY UNDERSTANDING PIPELINE FAILURE DRIVERS

Dilusha Weeraddana [1], Bin Liang [1,3], Zhidong Li [1,3], Yang Wang [1,3], Fang Chen [1,3],
Livia Bonazzi [2], Dean Phillips [2], Nitin Saxena [2],

[1]. Data61, CSIRO, Eveleigh, NSW, Australia
[2]. Western Water, Sunbury, Victoria, Australia
[3]. FEIT, University of Technology Sydney, NSW, Australia

ABSTRACT

Data61 and Western Water worked collaboratively to apply engineering expertise and Machine Learning tools to find a cost-effective solution to the pipe failure problem in the region west of Melbourne, where on average 400 water main failures occur per year. To achieve this objective, we constructed a detailed picture and understanding of the behaviour of the water pipe network by 1) discovering the underlying drivers of water main breaks, and 2) developing a Machine Learning system to assess and predict the failure likelihood of water main breaking using historical failure records, descriptors of pipes, and other environmental factors. The ensuing results open up an avenue for Western Water to identify the priority of pipe renewals.

CHALLENGES AND HIGHLIGHTS OF THE WORK

- While there is significant existing literature on pipeline failure causes, discovery of major failure factors was critical to discern which of these causes were the most important for Western Water;
- Thus, an in-depth analysis was carried out to identify underline pipe failure factors by data pre-processing through a sequence of steps;
- A Machine Learning prediction model was developed to identify future pipe failure likelihoods for every water main asset. These predictions were validated by separating the data into training and testing samples. Based on the prediction model, a derived list was generated and evaluated on the testing data;
- Some divergent trends were observed in the Western Water records (ex: failure rate for AC pipes decreases with the age). Therefore, Data mining techniques were used to explore the intricate interplay between age and other factors to reflect the true trend, and;
- Finally, a long-term forecasting model was developed for predicting which pipe assets are most likely to have a water main failure within the next twenty years. Furthermore, burst and fitting failures were considered separately;
- A user-friendly, end to end runnable tool was developed for the prediction.

INTRODUCTION

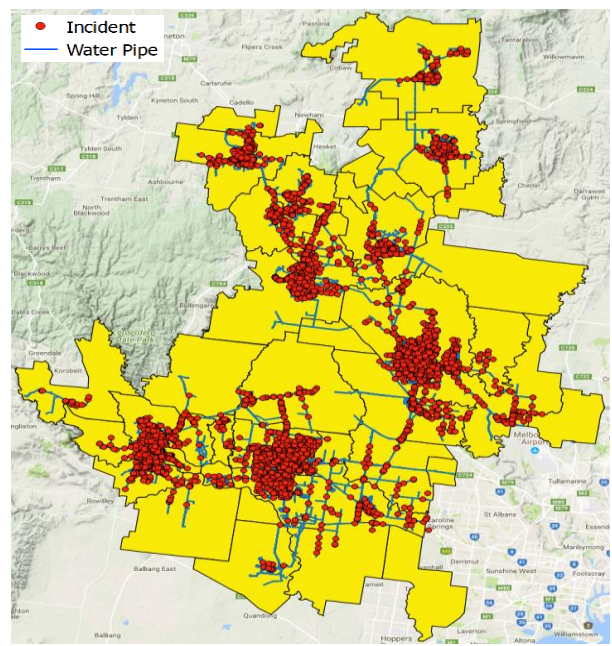

*Figure 1: Water main network of Western Water.*

The consequences of water pipeline failures can be extremely severe in-terms of water supply disruption, high repair cost and compensation claims. However, prediction of the water main breaks is not an easy task due to their low failure rate and high cost of inspection, which have led to a sparse historical data.

Data scientists at Data61 and engineering experts in Western Water commenced model development to answer these questions and ultimately produced more targeted break mitigation and asset renewal programs for Western Water. Western Water is one of Victoria's thirteen regional urban water corporations servicing 69,371 properties over an area of 3,000 square kilometres and a population of 160,339.

Mitigation of the water main breakage and water asset renewal programs should balance the consequence of water main failure and the cost to customers. To effectively achieve these dual objectives it is important to know: what are the causes of pipeline failure, what is the probability of a failure for an individual pipeline asset and the risk of these failures associated with the business?

Therefore, our main aim in this paper is to construct a detailed picture of factors affecting pipe failure rate, and predict the future pipe breakage likelihoods. These likelihoods will be used to calculate the risk distribution by combining with asset consequence factor data (Risk = Likelihood x Consequence), to develop a risk based investment decision framework for capital interventions.

Although the factors affecting pipe failures have been studied before, understanding of these factors is to a large extent incomplete due to their high complexity. Thus, comprehensive analyses were performed to identify the factors that lead to failures of water pipes. This involved exploring statistically significant correlation between water main breaks and operational factors sourced from Western Water's internal databases as well as external datasets such as the Bureaus of Statistics and Meteorology. In addition, a Machine Learning-based data analytic model was developed to predict the likely probabilities of future pipe failures. Data mining techniques were used to explore the intricate interplay between age and other factors to reflect the true trend of failure rate over the time. Finally, the probability of failure for an individual pipeline was calculated by extrapolating past performance of similar assets in similar operational conditions elsewhere in the network. An annual failure probability value was calculated for all water main assets until 2037.

The results were validated by comparing the number and location (suburb) of breaks projected by the modelling with actual performance in calendar year 2017. Validation was also carried out at the asset level by comparing assets with high failure probability against the asset renewal program. Further to this, end-to-end data analytic process is automated within Docker engine for the end user's convenience.

## EXISTING METHODS

Analysis of water pipe breakage and forecasting future failure rates has been studied over past few decades using variety of methods and frameworks.

Uri et al. (1979) developed a forecasting technique to study how the number of breaks would change with time if the pipes were not replaced. In that study, authors used a Poisson model based on the age of the pipes. Moreover, prediction of water main breaks has been studied using survival-based methods, such as Poisson regression by Asnaashari et al. (2009), and Weibull model.

Most recently, tree-based Machine Learning techniques have been used to analyse water pipe breakages in Syracuse, USA by Avishek et al. (2017) and in Queensland, Australia by Liang et al. (2017).

Although there is significant existing literature, there still exist open questions regarding intricate relationship among the major factors causing pipe failure, and their long-term effect on the life-time of a pipe. Thus, discovery of major failure factors is critical to discern which of these factors are the most important for different water utilities.

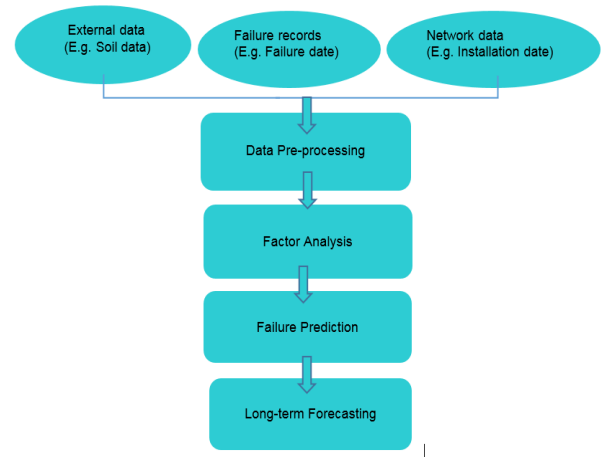

*Figure 2: Schematic for Data 61 data analytic model for failure prediction.*

## DATA ANALYTIC MODEL FOR FAILURE PREDICTION

The framework of the proposed model is depicted in Figure 2. The first step entails pre-processing pipe attribute data, and pipe failure data obtained from the Western Water's internal database. In the next step, the influential and significant factors are investigated and a water main failure prediction model is developed using a Machine Learning model (Random Forest Regression). Then the performance of the model is evaluated. Finally, a long-term failure forecasting model is developed, with an end-to-end runnable tool to automate the entire prediction process. The following subsections discuss the aforementioned processes.

**Data pre-processing**

There are three main data sources used as the input to the analytical model:

1) **Network data** describes water main information such as asset number, installation date, material, diameter, length, and location.
2) **Work order data** describes water main failure information such as asset number, failure date, location, and failure type (burst, fitting).
3) **External data** includes information in addition to assets, such as weather data from Bureau of Meteorology and census data from Australian Bureau of Statistics.

The above data should be sufficiently accurate for the intended use, so a data quality review has been undertaken based on three key characteristics: completeness, validation, and consistency (examination for invalid values). The quality review demonstrates that the data is sufficient and accurate for further analysis. Accordingly, this process allows to establish a comprehensive data file with complete information for each asset that can be used as an important input to further analysis.

Moreover, when information is gathered from multiple sources, and prior to adoption of advanced analytic techniques it is essential to match the failure records with the network data and identify gaps in the datasets. In addition, environmental and demographic factors need to be matched with the network data. Specifically, failure records and information are assigned to the corresponding assets based on the work order number, and environmental and demographic information are assigned to the assets based on the geographic locations.

**Factor analysis**

Factor analysis has been used to identify pipeline failure drivers and compare their relative impact on the network based on the water network information.

Factor analysis measures the correlation between asset performance based on the comprehensive data and a large range of factors (including environmental, demographic, asset specific factors). While there is significant existing literature on pipeline failure causes, this step is critical to discerning which of these causes would be the most important for Western Water. The asset performance is based on failure rate which is the number of asset failures per 100km per year. Both single factor analysis and multi-factor analysis have been performed to identify the possible driving factors. The asset performance usually is not related to only one factor, so it is essential to measure the correlation based on multiple factors. Compared to the single factor analysis, multi-factor analysis is a factorial method devoted to study a group of individuals which is described by a set of factors.

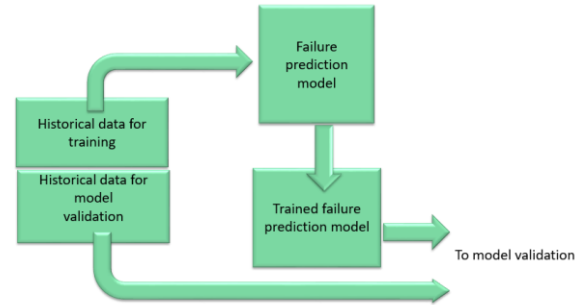

*Figure 3: Pipe failure prediction model.*

**Pipeline failure prediction**

This phase involves predicting future (short-term) water pipe failure probabilities. We framed this scenario as determining the likelihood of failure on each given pipe within the next immediate years.

The model we developed includes specialized algorithms to handle the large amount of numerical calculations and data for prediction. The underlying statistical principle employed here is the Random Forest Regression, as trees are ideal candidates to capture complex interaction in the pipe data. This model is initially reported by Breiman (2001) and extended by Harvey et al. (2014). Random Forest Regression captures and extrapolates non-linear interactions among failure factors.

The failure prediction is generated by training the Machine Learning model on historical failure records and other factors. Prediction accuracy is achieved by running many iterations of non-linear regression and then averaging the results. Finally, this trained model produces a failure probability score for each water main asset. This process is schematically illustrated in Figure 3.

**Long-term forecasting**

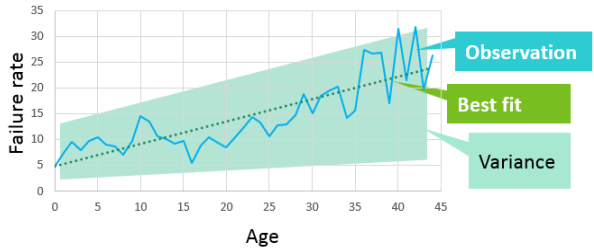

*Figure 4: Failure rate of AC pipes with the increase of the pipe age.*

This phase extends the short-term prediction results to forecast pipe failures 20 years into the future. Here we assume that the function of failure rate with age is linear. Which means the failure rate will increase with constant value for each year.

Our approach on choosing the optimal coefficient, $a^*$ for the trend shown in Figure 4 is given below:

1. Calculate the maximum and minimum values of the coefficients from single data point $(A_i, FR_i)$, which are:

$$a_{min} = min\{\frac{FR_i}{A_i}\}_i, a_{max} = max\{\frac{FR_i}{A_i}\}_i$$

2. Set the step-size as $\epsilon$. For integers $k$ ($k: a_{min} \leq a_{min} + k \cdot \epsilon \leq a_{max}$), obtain the sequence $\{L_k\}$ of sum of squared regression errors (loss), each of which is calculated as $\sum_i(FR_i - (a_{min} + k \cdot \epsilon) \cdot t)^2$

3. Select the smallest value of $\{L_k\}$, return the corresponding $a^*$ as optimal.

Where, $A_i, FR_i$ are age and failure rate of each pipe, $i$. To obtain the data points shown in Figure 5, we must use a set of pipes (this is because individual pipe can only provide a small number of points with high variance). Here, the set of pipes can be a category. The category was initially fixed manually, e.g. all AC pipes. Thus, an optimum coefficient is calculated for each pipe type.

RESULT ANALYSIS

**Data pre-processing**
First of all, a data pre-processing task was carried out to clean the raw data and match pipe attribute data with failure records. Data cleaning was conducted to make sure the data is complete and valid.

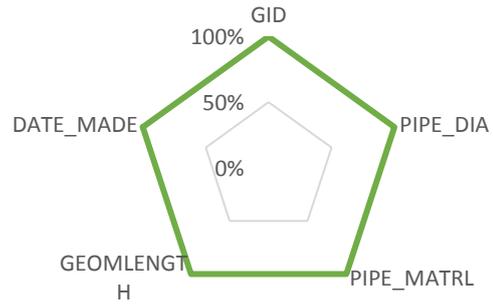

*Figure 5: Analysis for the completeness of water pipes.*

- **Completeness**: this is a statistic that does not allow empty values. For water pipes, all the records are complete, as shown in Figure 5. However, for failure incident records, 990 records have empty EVENT_DATE values, with 95% of completeness.

- **Validity**: this is a statistic that does not allow invalid values. For water pipe data, 3432 records have invalid DATE_MADE values, making 98.5% validity (see Figure 6). Failure data include 50 invalid records making it 99% valid for processing.

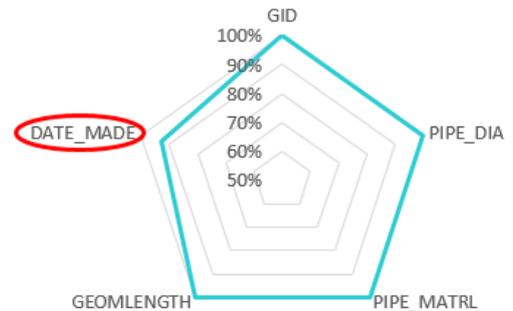

*Figure 6: Analysis for the validity of water pipes.*

Figure 7 shows the data matching process, which matches the network data with work order data. Over 90% of data can be successfully matched.

*Table 1: Overall data matching process*

| All Records | 93.56% |
|---|---|
| Bursts | 93.53% |
| Fitting Failures | 93.57% |

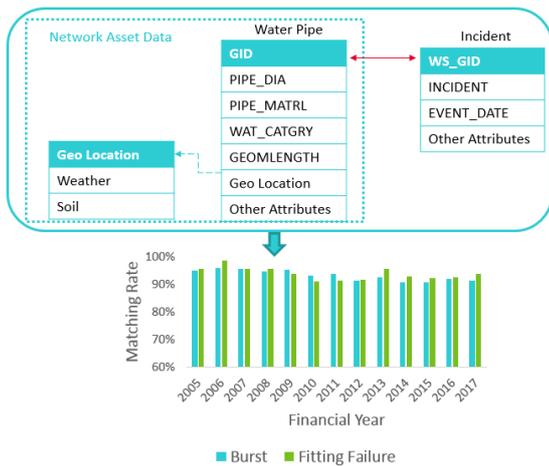

*Figure 7: Data matching: Process of matching failure records with pipe attributes.*

Ultimately, the data pre-processing including the quality review has demonstrated that the data is sufficient and accurate for further analysis. Using the processed data, overall failure rates for burst and fitting failures were calculated for each year from 2005 to 2016, as depicted in Figure 8.

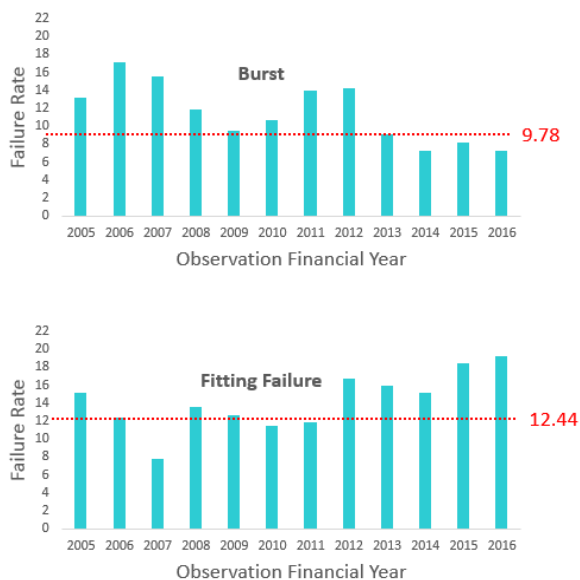

*Figure 8: The failure rate of failure type for financial year 2005-2016.*

**Factor analysis outcomes**
Factor analysis allowed Western Water to compare the relative impact each factor has on causing failures. For example, within operational factors, AC mains were found to failure more often than others (See Figure 9). It was also found that water mains with laid year before 1985 exhibit higher failure rates. (See Figure 10).

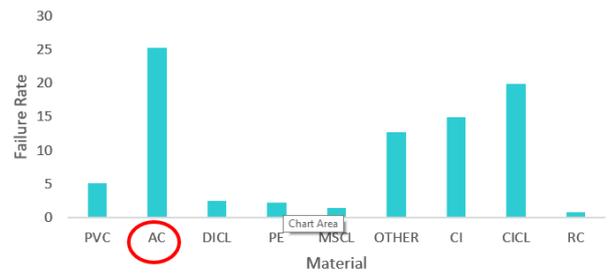

*Figure 9: Failure rate of water mains based on materials.*

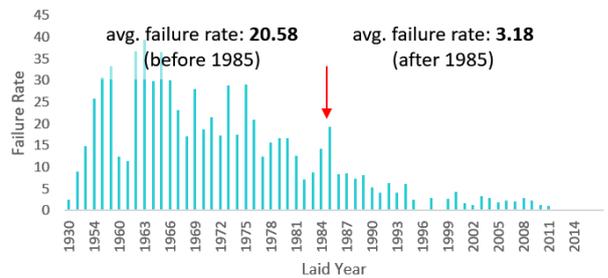

*Figure 10: Failure rate of water mains based on laid years.*

Furthermore, environmental factors have been analysed, including weather data and soil data. The weather data is extracted from Bureau of Meteorology. Monthly mean temperature data over ten years (2006~2015) was used for the analysis. The analysis results show that with the increase of temperature, overall failure rates increase (See Figure 11).

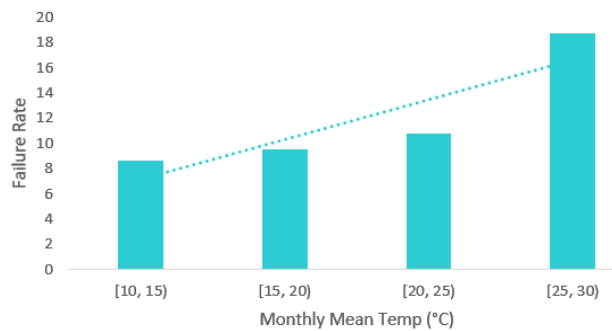

*Figure 11: Failure rate of water mains based on monthly mean temperatures.*

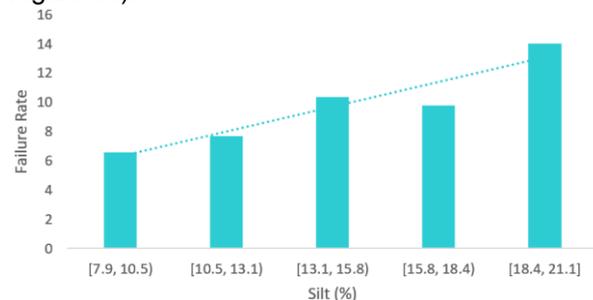

*Figure 12: Failure rate of water mains based on silt values.*

The soil data is from Soil and Landscape Grid[1]. Five types of soil data were available for analysis: clay, bulk density, silt, sand, and pH. It can be seen in Figure 12 that with the increase of silt value (%), the overall failure rates increase. However, less significant correlation was found between water main failures and other soil types.

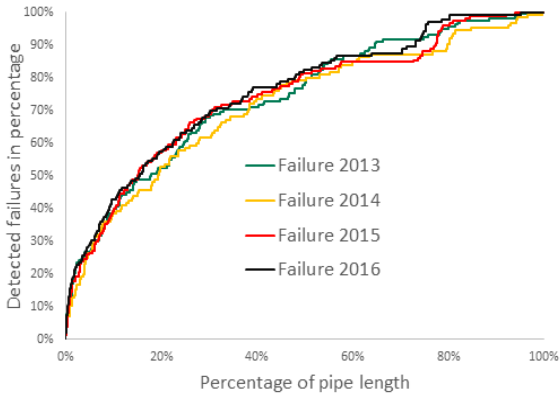

Figure 13: Validation of the prediction model based on historical data. Ex: For burst failure, model was trained from 2005-2012, 2005-2013, 2005-2014, 2005-2015 and tested on 2013,2014,2015,2016 data.

**Failure prediction outcomes**
Pipe length based:

Firstly, the prediction model was calibrated using the failure records from 2005-2015. Then the calibrated model was applied to predict the pipe failure probability for each pipe from year 2013 to 2016. Then the pipes were ranked according to the failure probability of each pipe. Using the ranked list, actual breaks from highest probability to lowest probability are accumulated (cumulative sum of breaks). The percentage of detected breaks is plotted against the percentage of inspected pipe lengths. Figure 13 shows that, if the first 10% pipes are inspected, more than 30% of burst failures can be detected.

Suburb based:

Results were also validated based on suburb. Here, the suburbs were ranked according to the accumulated failure probability of each pipe in a suburb. Using the ranked list, actual breaks from highest probability to lowset probability are accumulated (cumulative sum of breaks). The percentage of detected breaks is plotted against the percentage of inspected suburbs. Figure 14 shows that, if the first 10 suburbs are inspected, more than 70% of burst failures can be detected.

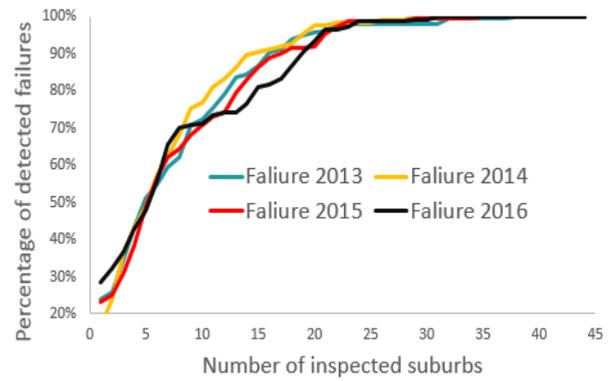

Figure 14: Suburb level model validation for burst failures.

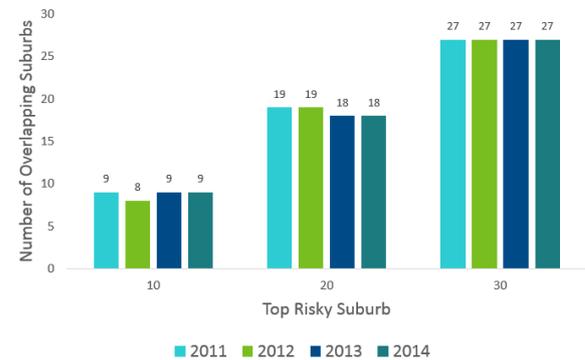

Figure 15: Number of overlapping suburbs between model output and actual burst failure.

Moreover, if the top 30 suburbs are inspected, 27 overlapping suburbs can be found based on our model, as illustrated in Figure 15.

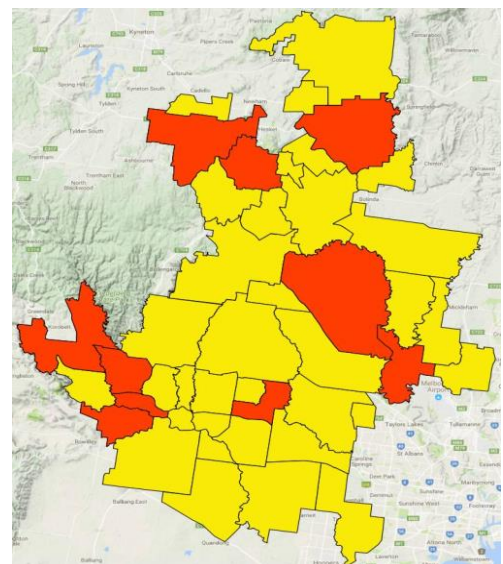

Figure 16: Model output for top risking fitting failure zones for 2016.

---

[1] http://www.clw.csiro.au/aclep/soilandlandscapegrid/GetData-DAP.html

*Table 2: Top 10 risky suburbs for fitting failures*

| 2013 | 2014 | 2015 | 2016 |
|---|---|---|---|
| Bacchus Marsh | Bacchus Marsh | Bacchus Marsh | Bacchus Marsh |
| Sunbury | Sunbury | Sunbury | Kurunjang |
| Darley | Melton | Melton west | Darley |
| Melton | Gisborne | Woodend | Melton west |
| Brookfield | Darley | Kurunjang | Woodend |
| Woodend | Brookfield | Gisborne | Melton |
| Maddingley | Wildwood | Darley | Sunbury |
| Melton west | Kurunjang | Melton | Maddingley |
| Gisborne | Woodend | Brookfield | Brookfield |
| Kurunjang | Melton west | Diggers rest | Gisborne |

Suburbs are sorted by their normalized fitting failure probability in descending order. The top 10 risky suburbs from 2013-2016 have been listed in the Table 2. The suburbs highlighted in green are overlapped with actual failure records. Hence, in each year, our model can successfully detect 9 out of top 10 risky suburbs accurately.

Figure 16 shows the top 10 risky suburbs for fitting failure in 2016.

Long-term burst failure prediction for all water main assets is given in Figure 17. Note that the prediction is a probability distribution where darker areas represent high probability estimates and the lightly shaded upper and lower bounds represent low probability estimates. The uncertainty arises due to statistical significance of correlations from factor analysis.

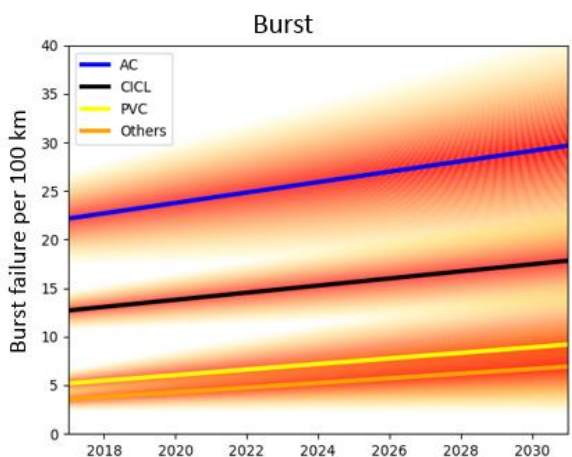

*Figure 17: Projected water main fitting failure for the Western Water network over the next 15 years. Failure prediction is calculated as a probability distribution, darker shades of red represent a higher probability to occurrence.*

When we consider the entire spectrum of the years, failure rate increases with the age (see Figure 4). In recognition of this pattern, the failure rate gradually rises over the years starting from 2017. Thus, the mean prediction or the trajectory most likely to occur is represented with a solid line in Figure 17 and Figure 18. Our model predicts that by 2030 there will be an increase of 22%, 26% in the burst and fitting failures respectively.

**Significance of our approach:**

- Water Mains in more than 40 suburbs in Victoria are studied.
- Data was pre-processed for application through sequence of steps.
- The factors were analysed for their impact to failures.
- Prediction model was developed and evaluated on historical data.
- Evaluation shows more than 20-40% failures can be detected by inspecting 10% of total pipe length.
- Failure likelihood of each pipe for next 20 years was predicted, and further analysed based on material and pipe size.
- Failure prediction tool was developed, which automates the process from data cleaning to long-term forecasting, as illustrated in Figure 19.

Our model provides a projection of the likelihood of pipe failure. These likelihoods along with the consequence of failures are being used in Western Water's current investment planning, to make risk based investment decisions for capital interventions. The severity of the consequence of failures is determined with the input from Western Water's internal data.

CONCLUSION

This project has been developed to provide assistance to forecast and plan water main renewals with more confidence via predictive analytics. Pipeline maintenance and renewal programs balance level of service requirements and the need to minimize cost to customers. Therefore, we constructed a complete picture of factors causing pipe failures in Western Water's water pipeline network, and developed a prediction model to estimate the probabilities of water main breaks based on those factors. Results demonstrate that our model is capable to provide valuable assistance to forecast and plan water main renewals with more confidence via predictive analytics. The next step is to apply a consequence rating to enhance the model predictions, as both of these factors are important to identify the priority of pipe renewals. Ultimately, we believe this work, at the intersection of Machine Learning and Asset Management, will lead to more effective and proactive infrastructure maintenance in the Australian water industry.

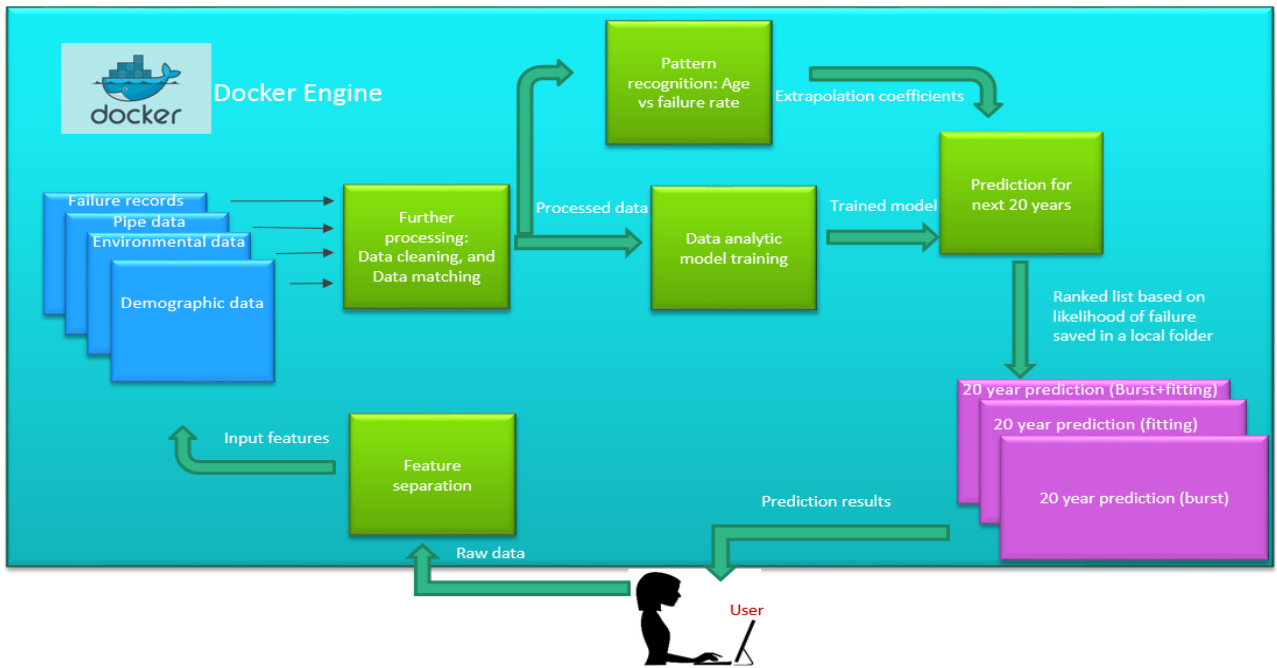

*Figure 19: Schematics for the end to end process automated within Docker engine for ease use for the end users. User only needs to provide the input data files. System will go through several steps ranging from feature extraction, data cleaning, data matching, and pattern recognition to failure forecasting. The output files (ranked list based on likelihood of failure) will be saved in a separate folder. Burst, fitting and overall failures are considered as separate cases and three output files will be generated respectively.*

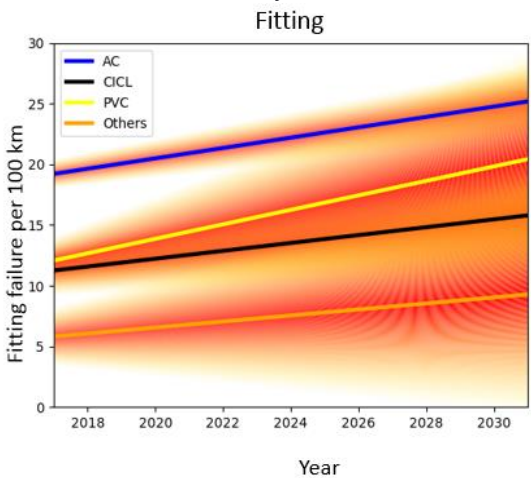

*Figure 18: Projected water main fitting failure for the Western Water network over the next 15 years. Failure prediction is calculated as a probability distribution, darker shades of red represent a higher probability to occurrence.*

## COPYRIGHT